	\documentclass[final,dvips]{aipproc}

\layoutstyle{8x11double}

\usepackage{amsmath}
\usepackage{amssymb}
\usepackage{epsfig}
\usepackage{graphicx}
\usepackage{stmaryrd}
\usepackage[dvips]{color}
\def\10{$SO(10)$}
\def\21{SU(2) $\otimes$ U(1) }

\def\422{$SU(4) \otimes SU(2) \otimes SU(2)$}
\def\321{SU(3) $\otimes$ SU(2) $\otimes$ U(1)}
\def\lsim{\raise0.3ex\hbox{$\;<$\kern-0.75em\raise-1.1ex\hbox{$\sim\;$}}}
\def\gsim{\raise0.3ex\hbox{$\;>$\kern-0.75em\raise-1.1ex\hbox{$\sim\;$}}}

\usepackage{dcolumn}
\usepackage{bm}
\usepackage{epsfig}

\newcommand {\ignore}[1]{}

\newcommand{\be}{\begin{equation}}
\newcommand{\ee}{\end{equation}}
\newcommand{\bea}{\begin{eqnarray}}
\newcommand{\eea}{\end{eqnarray}}

\newcommand{\eV}{\mathrm{eV}}
\newcommand{\keV}{\mathrm{keV}}

\newcommand{\GeV}{\mathrm{GeV}}
\def\baselinestretch{1.0}
\def\roughly#1{\mathrel{\raise.3ex\hbox{$#1$\kern-.75em
      \lower1ex\hbox{$\sim$}}}} \def\lsim{\roughly&lt;}
\def\gsim{\roughly&gt;}
\def\e6{E(6)}
\def\321{$SU(3)_{c}\otimes SU(2)_L \otimes U(1)$}
\def\10{SO(10)}

\def\422{SU(4) $\otimes$ SU(2) $\otimes$ SU(2)}

\newcommand{\Gyr}{\mathrm{Gyr}}

\newcommand{\GG}{\Gamma_{18}}

\newcommand{\oxf}{Oxford Astrophysics, Denis Wilkinson 
Building, Keble Road, OX1 3RH, Oxford, UK }

\begin{document}

\title{Decaying majoron dark matter and neutrino masses}


\author{Massimiliano Lattanzi}{
  address={\oxf}
}





\begin{abstract} 
 
We review the recent proposal by Lattanzi \& Valle of the majoron as a suitable warm 
dark matter candidate. The majoron
is the Goldstone boson associated to the spontaneous breaking of ungauged lepton
number, one of the mechanisms proposed to give rise to neutrino masses. 
The majoron can acquire a mass through quantum gravity effects, and can 
possibly account for the observed dark matter component of the Universe.
We present constraints on the majoron lifetime, mass and abundance obtained
by the analysis of the cosmic microwave background data. We find that, in the 
case of thermal production, the limits for the majoron mass read $0.12~\keV~<~m_J<~0.17~\keV$, 
and discuss how these limits are modified in the non-thermal case.
The majoron lifetime $\tau$ is constrained to be $\gtrsim 250\,\Gyr$.. 
We also apply this results to a given seesaw model for the generation of neutrino masses, and
find that this constraints the energy scale for the lepton number breaking phase transition
to be $\gtrsim 10^6\,\GeV$.
We thus find that the majoron decaying dark matter (DDM) scenario fits nicely in models where neutrino masses
  arise {\it a la seesaw}, and may lead to other possible cosmological
  implications.  
\end{abstract}

\maketitle


\subsection{Introduction}

Understanding the nature of dark matter (DM) and its origin represents one of the
longest-standing challenges in particle cosmology. We know from cosmological
observations \cite{Spergel:2006hy, Tegmark:2006az} that only $\sim 5\%$ of the Universe energy content is accounted for
by normal, baryonic matter, while the remaining is in the form of dark matter 
($\sim 25\%$) and of a similarly elusive energy component, dubbed dark energy 
($\sim 70\%$).

Historically, the neutrino was at first seen as the natural dark matter candidate, due
to its weak interaction with ordinary matter \cite{Cowsik:1972gh}.
However, it soon became evident that the high velocity dispersion of the relativistic neutrinos would erase 
all density perturbations below a critical scale of some tens of megaparsecs {\cite{Bond:1980ha}}, 
thus completely spoiling the whole process of structure formation. This critical scale
is called free-streaming length; dark matter candidates with a large free 
streaming length, like the neutrino, are classified as Hot Dark Matter (HDM). 
Nowadays, although we know from neutrino oscillation experiments that
neutrinos do have mass~\cite{Maltoni:2004ei}, recent cosmological
data~\cite{Lesgourgues:2006nd} as well as searches for distortions in
beta~\cite{Drexlin:2005zt} and double beta decay
spectra~\cite{Klapdor-Kleingrothaus:2004wj} place a stringent limit on
the absolute scale of neutrino mass that precludes neutrinos from
being viable dark matter candidates~\cite{Gelmini:1984pe} and
from playing a {\sl direct} role in structure formation.

Many candidates for the dark matter particle are presently under consideration: 
among the most popular, the supersymmetric 
neutralino, and the Kaluza-Klein particles (see \cite{Bertone:2004pz} and references therein). 
Most of this candidates share the 
property of being Cold Dark Matter (CDM) particles, because their velocity dispersion,
and consequently their free streaming length, are so small to be practically irrelevant
for cosmological structure formation. 
This avoids the problem of small scale damping of HDM models, and in fact CDM models agree well with
observations down to scales of several Mpc, once mildly non-linear
effects are taken into account \cite{Tegmark:2006az}.
 However, it seems that
the CDM scenario is unable to reproduce the matter distribution at the smallest
scales, i.e, on Mpc scales and below (see \cite{Ostriker:2003qj} and references
therein). First, it predicts a number of dwarf galaxies
much larger than observed. Secondly, numerical simulations produce DM halos with very high
density cores, but this cuspiness is not actually observed in real galactic cores.
It is still unclear if these are
shortcomings of the model itself, or instead come from our poor understanding of 
astrophysical processes important at the scale of interest, or
even from numerical issues related to the high non-linearity of the phenomena under consideration.
The problem with the CDM scenario is in some sense opposite 
with respect to the HDM one: where the latter predicts too little power in the 
small scale fluctuations, the former predicts too much of it. In other words: a HDM Universe
is too smooth with respect to the observed one, while a CDM Universe is too clumpy.

Between the two limiting cases of hot and cold dark matter, lies the so-called 
Warm Dark Matter (WDM). Examples of WMD candidates include the sterile 
neutrino \cite{Dodelson:1993je} and the light gravitino \cite{Pagels:1981ke}. 
The free streaming length of WDM particles is 
in the Mpc range, thus quite smaller with respect to the typical HDM value (hence the name). 
This is appealing because it suggest the possibility of keeping the successful predictions predictions
of the CDM scenario at the intermediate and large scales, and at the same time alleviating (and hopefully eliminating)
the small-scale inconsistencies of the model \cite{Bode:2000gq}. 
Here we describe our recent proposal of a WDM model linking the problem of dark matter with the issue
of the origin of neutrino masses \cite{Lattanzi:2007ux}.
\subsection{Majoron dark matter}
If neutrino masses arise from the spontaneous violation of ungauged
lepton number there must exist a pseudoscalar gauge singlet
Nambu-Goldstone boson, the
majoron~\cite{chikashige:1981ui,schechter:1982cv}. This may pick up a
mass from non-perturbative gravitational effects that explicitly break
global symmetries~\cite{Coleman:1988tj}. Despite the fact that the majorons 
produced at the corresponding
spontaneous L--violation phase will decay, mainly to neutrinos, they
could still provide a sizeable fraction of the dark matter in the
Universe since their couplings are rather tiny.

This scenario was first considered in Ref. ~\cite{Berezinsky:1993fm}; 
however, since then there have
been important observational developments which must be taken into
account in order to assess its viability, most notably the recent
 cosmological microwave observations from the Wilkinson Microwave Anisotropy Probe
(WMAP)~\cite{Spergel:2006hy}. 

%
%

%
%

%

In the following we then consider the majoron decaying dark matter (DDM)
idea in a modified $\Lambda$CDM cosmological model in which the dark
matter particle is identified with the weakly interacting majoron $J$
with mass in the keV range. A keV weakly interacting particle could provide a sizeable fraction of
the critical density $\rho_{cr} = 1.88 \times 10^{-29} h^{2}$
$\mathrm{g/cm^{3}}$ and possibly play an important role in structure formation,
since the associated Jeans mass $m_{Jeans} \sim m_{Pl}^3/m_J^2$ lies in the relevant
range.

The majoron is however not stable but decays non
radiatively with a small decay rate $\Gamma$.  In this DDM scenario,
the anisotropies of the cosmic microwave background (CMB) can be used
to constrain the lifetime $\tau=\Gamma^{-1}$ and the present abundance
$\Omega_{J}$ of the majoron; here we show that the cosmological
constraints on DDM majorons not only can be fulfilled but also can
easily fit into a comprehensive global picture for neutrino mass
generation with spontaneous violation of lepton number. 
For definiteness here we adopt the very
popular possibility that neutrino masses arise {\sl a la
  seesaw}~\cite{Valle:2006vb}.

\paragraph{Majoron abundance} 
Although majorons could result from a phase transition, we first
consider them to be produced thermally, in equilibrium with photons in
the early Universe. In this case the majoron abundance $n_J$ at the
present time $t_0$ will be, owing to entropy conservation and taking
into account their finite lifetime:
\begin{equation}
\frac{n_J(t_0)}{n_\gamma(t_0)}=
\frac{43/11}{N_D}\frac{n_J(t_D)}{n_\gamma(t_D)}\;e^{-t_0/\tau},
\end{equation}
where $t_D$ is the time of majoron decoupling, and $N_D$ denotes the
number of quantum degrees of freedom at that time. The exponential factor
accounts for majoron decay. If $T(t_D)\gtrsim
170\,\GeV$, then $N_D=427/4=106.75$ for the particle content of the standard
model, while in the context of a supersymmetric extension
of the SM, there would possibly be, at sufficiently early times, about
twice that number of degrees of freedom. Finally, 
in thermal equilibrium the majoron to photon ratio $f \equiv
n_J(t_D)/n_\gamma(t_D)$ is equal to 1/2.
The present density parameter of majorons is then, using $N_D=106.75$:
\begin{equation}
\Omega_J h^2=\frac{m_J}{1.25\,\keV}e^{-t_0/\tau}.
\end{equation}

Another possibility is that majorons were produced already out of equilibrium.
In this case there is a range of possible models, that we can write generically as
\begin{equation}
\Omega_J h^2=\beta \frac{m_J}{1.25\,\keV}e^{-t_0/\tau}.
\label{eq:omj}
\end{equation}
where the quantity $\beta$ parametrizes our ignorance about both the
exact production mechanism, and the exact value of $N_D$. When $\beta
=1$, we recover the scenario described above, with $f	=1/2$ and
$N_D=427/4$.

\begin{center}
\begin{figure}[h]
\includegraphics[clip, width=\linewidth]{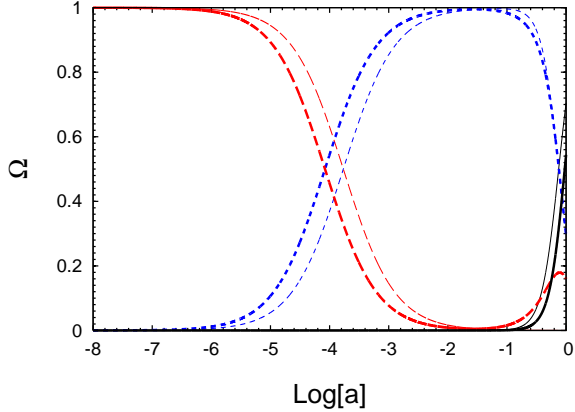}
\caption{(Color online) Evolution of abundances in the standard (thin lines) and DDM
  (thick lines) universe scenario: blue/short dashed, red/long dashed and black/solid correspond to the
  matter, radiation and $\Lambda$ components, respectively.}
\label{fig:lss}
\end{figure}
\end{center}

\paragraph{Effects of majoron DM on the CMB}
Clearly if the majoron has to survive as a dark matter particle it must
be long-lived, $\tau \geq t_0$. However, a more stringent bound
follows by studying the effect of a finite majoron lifetime on the
cosmological evolution and in particular on the CMB anisotropy
spectrum. In the DDM scenario, due to particle decays, the dark matter
density is decreasing faster than in the standard cosmological
picture.  This changes the time $t_{eq}$ of radiation-matter equality.
This means that, for a fixed $\Omega_J$, there will be more dark
matter at early times, and the equality will take place earlier, as
illustrated in Fig.~\ref{fig:lss}.
The present amount of dark matter is $\Omega_{DM} = 0.25$ for both
models; $\Gamma^{-1} = 14\,\Gyr$ in the DDM model. Other relevant
parameters are $\omega_b = 2.23\times 10^{-2}$ and $h=0.7$. The time
at which the blue and red lines cross is the time of matter-radiation
equality; for fixed $\Omega_{DM}$, it shifts to earlier times as the
majoron lifetime decreases.


The time of matter-radiation equality has a direct effect on the CMB
power spectrum. The gravitational potentials are decaying during the
radiation dominated era; this means that photons will receive an
energy boost after crossing potential wells. This so-called early
integrated Sachs-Wolfe (EISW) effect ceases when matter comes to
dominate the Universe, since the potential are constant during matter
domination. The overall effect is to increase the power around the
first peak of the spectrum as the equality moves to later times.

On the other hand, since $\tau \gtrsim t_0$, we expect the majoron
decays to make the gravitational potentials vary again in the late
stage of the cosmological evolution. This will induce a similar effect
to the one described above, only affecting larger scales due to the
increased horizon size. This late integrated Sachs-Wolfe (LISW) effect
results then in an excess of power at small multipoles.


Both effects can be used in principle to constrain the majoron
lifetime and cosmological abundance. In order to carry on a quantitative
analysis, we
have developed a modified version of the \verb+CAMB+
code~\cite{Lewis:1999bs}, that enables to compute the CMB
anisotropy spectrum once the majoron lifetime and abundance are given
in addition to the standard $\Lambda$CDM model parameters.

We stress the fact that even if a keV majoron constitutes a warm dark matter particle, it
actually behaves as cold dark matter insofar as the calculation of its effect
on the CMB spectrum is concerned, since CMB measurements cannot
discriminate between cold and warm dark matter. The
latter behaves differently from a cold one on scales smaller than its
free-streaming length $\lambda_{fs}$. For a particle mass in the keV
range, we have $\lambda_{fs}\sim 1\,\mathrm{Mpc}$ which corresponds in
the CMB to a multipole $\ell\sim \textrm{few thousands}$.

The formalism needed to account for the cosmological evolution of an
unstable relic and of its light decay products, has been developed for
example in Refs.  \cite{Kaplinghat:1999xy,Ichiki:2004vi}, including the
modifications in both background quantities and perturbation
evolution. We report here the necessary changes to the evolution equations, 
using the formalism introduced in Ref. \cite{Ma:1995ey}. The subscript $J$ denotes
the majoron dark matter component, while the subscript $DP$ denotes
the majoron relativistic decay products.

\subsubsection{Background equations}
The equations for the time evolution of the dark matter and decay 
products energy density are:

\begin{subequations}
\begin{align}
&\dot\rho_J+3\frac{\dot a}{a}\rho_J=-a \Gamma \rho_J, \\
&\dot\rho_{DP}+4\frac{\dot a}{a}\rho_{DP}=a \Gamma \rho_J,
\end{align}
\end{subequations}
where $a$ is the cosmological scale factor, and the dot denotes
derivative with respect to conformal time (hence the extra $a$ factor on
the right hand side). Note that the source term for the decay
products involves the dark matter density, thus effectively coupling
the two equations.

\subsubsection{Perturbation equations}

The perturbations in the DM and DP components evolve according to the following
set of equations (see Ref. \cite{Ma:1995ey} for the meaning of the symbols):

\paragraph{Majoron dark matter}
\begin{equation}
\dot{\delta_{J}}=-\frac{\dot h}{2}
\end{equation}

\paragraph{Decay products}
\begin{subequations}
\begin{align}
\dot\delta_{DP} &= -\frac{2}{3}\left(\dot h+2\theta_{DP}\right)+\frac{\dot r}{r}\left(\delta_J-\delta_{DP}\right) \label {eq:delta_DP}\\
\dot\theta_{DP} &= k^2\left(\frac{\delta_{DP}}{4}-\sigma_{DP}\right)-\frac{\dot r}{r}\theta_{DP}\\
\dot\sigma_{DP} &= \frac{2}{15}\left(2\theta_{DP}+\dot h+6\dot\eta\right)-\frac{3}{10}kF_{DP,3}-\sigma_{DP}\frac{\dot r}{r}\\
\begin{split}
\dot F_{DP,\ell} &= \frac{k}{2\ell+1}\left[\ell F_{DP,\ell-1}-(\ell+1)F_{DP,\ell+1}\right] \\
&-F_{DP,\ell}\frac{\dot r}{r} ,\qquad \ell\ge 3. 
\end{split}
\end{align}
\end{subequations}

where $r$ denotes the ratio of DP to photons energy densities, i.e. $r\equiv \rho_{DP}/\rho_{\gamma}$ (any fiducial density scaling as $a^{-4}$ will actually do the job). The presence of the dark matter density perturbation
$\delta_J$ in the right hand side of Eq. (\ref{eq:delta_DP}) again couples the two sets of equations.

\subsection{Statistical analysis}
\paragraph{Parametrization}
Two distinct mechanisms effective at very different times characterize
the effect of DDM on the CMB.
It is therefore convenient to choose a parametrization that can take
advantage of this fact.
%
In particular, the ``natural'' parametrization $(\Omega_J,\,\Gamma)$
has the drawback that both parameters affect the time of
matter-radiation equality.  It is more convenient to define the
quantity
\begin{equation}
Y \equiv \left.\frac{\rho_J}{\rho_b}\right|_{t=t_{\mathrm{early}}},
\end{equation}
where $\rho_b$ is the energy density of baryons, and
$t_{{\mathrm{early}}}\ll t_0 \lesssim \tau$.  As long as this
condition is fulfilled, the value of $Y$ does not depend on the
particular choice of $t_\mathrm{early}$, since the ratio
$\rho_J/\rho_b$ is asymptotically constant at small times. Given that
$t_{eq}\ll\tau$ we can use the value of $Y$ to parametrize the
relative abundance of majorons at matter-radiation equality.
In order to simplify notation let us also define $\GG \equiv
\Gamma/(10^{-18} \mathrm{sec}^{-1})$; in this way, $\GG=1$ corresponds
to a lifetime $\tau \simeq 30\, \Gyr$.

The advantage of using the parametrization $(Y,\, \Gamma)$ is that,
when all other parameters are fixed, the time of matter-radiation
equality is uniquely determined by $Y$, while the magnitude of the
LISW effect is largely determined by $\Gamma$.

\begin{center}
\begin{figure}[h]
\includegraphics[clip, height=8cm]{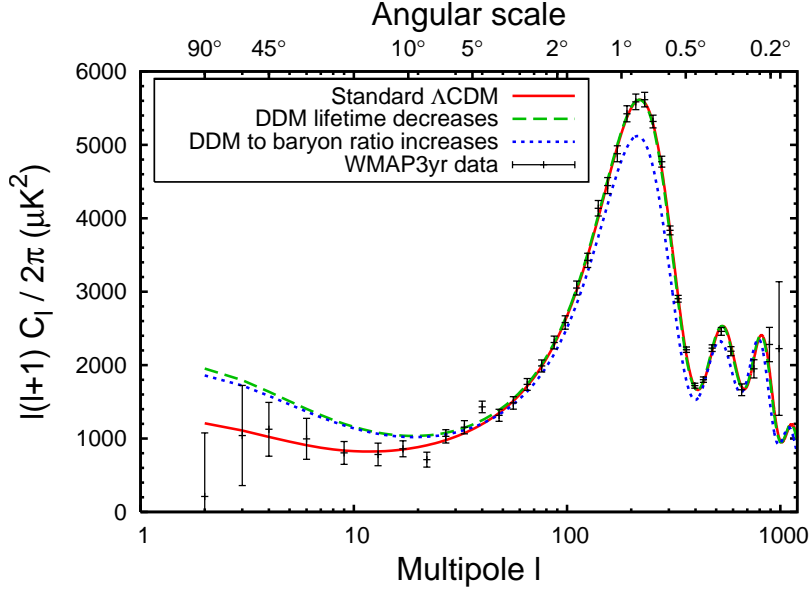}
\caption{(Color online) Effect of DDM parameters on the CMB anisotropy spectrum. The value of the parameters are as follows. Red/solid: fiducial model $(\GG,\,Y) = (0,\,4.7)$. 
Green/dashed: $(\GG,\,Y) = (1.2,\,4.7)$.
Blue/dotted: $(\GG,\,Y) = (1.2,\,5.6)$. See text.
}
\label{fig:cmb}
\end{figure}
\end{center}

We show in Fig.
\ref{fig:cmb} how the two physical effects are nicely separated in
this parametrization.
We start from a fiducial model with $\GG = 0$ and $Y = 4.7$; all other
parameters are fixed to their WMAP best-fit values.  The values of
$\GG$ and $Y$ are chosen in such a way to give $\Omega_J h^2 = 0.10$,
so that this fiducial model reproduces exactly the WMAP best-fit.
At a larger majoron decay rate $\GG = 1.2$, i.e., $\Gamma^{-1}\simeq
27\,\Gyr$, the LISW effect makes, as expected, the power at small
multipoles increase, while the shape of the spectrum around the first
peak does not change, since the abundance of matter at early times has
not changed.
Finally, increasing $Y$ by 20\% the height of the first peak decreases
accordingly, while the largest angular scales (small $\ell$s) are
nearly unaffected. A small decrease in power in this region is
actually observed, and can be explained by noticing that increasing
the matter content we delay the onset of the $\Lambda$ dominated era,
reducing the $\Lambda$ contribution to the LISW effect.

Another advantage of using the above parametrization is that
$Y$ is directly related to the majoron mass through:
\begin{equation}
Y = 0.71 \times \left(\frac{m_J}{\keV}\right)\,\left(\frac{\beta}{\Omega_b h^2}\right) .
\end{equation}
\subsubsection{Results and discussion}
We are now ready to compute the constraints that CMB observations put
on the majoron abundance and lifetime. 
As seen from Fig.  \ref{fig:cmb}, even a lifetime twice as long as the
present age of the Universe, is quite at variance with respect to the
WMAP data.  
However one must take into account the fact the values of the other
cosmological parameters can be arranged in such a way as to reduce or
even cancel the conflict with observation, i.~e. degeneracies may be
present in parameter space.  
In order to obtain reliable constraints for the majoron mass and
lifetime, we perform a statistical analysis allowing for the variation
of all parameters.  
This is better accomplished using a Markov chain Monte Carlo approach;
we used to this purpose the widely known \verb+COSMOMC+ code
\cite{Lewis:2002ah}.

In our modified flat ($\Omega = 1$) $\Lambda$CDM model, all the dark
matter is composed of majorons. This means that no stable cold dark
matter is present~\footnote{This happens e.~g. in models where
  supersymmetry with broken R parity is the origin of neutrino
  mass~\cite{Hirsch:2004he,Hirsch:2000ef}.}.
The 7-dimensional parameter space we explore therefore includes the
two parameters $(Y,\,\Gamma)$ defined above, in addition to the five
standard parameters, namely: the baryon density $\Omega_b h^2$, the
dimensionless Hubble constant $h$, the reionization optical depth
$\tau_{re}$, the amplitude $A_s$ and spectral index $n_s$ of the primordial
density fluctuations. 
The cosmological constant density $\Omega_\Lambda$ depends on the
values of the other parameters due to the flatness condition. We
compare our results with the CMB anisotropies observed by the WMAP
experiment.
Once the full probability distribution function for the seven base
parameters has been obtained in this way, the probability densities
for derived parameters, such as the majoron mass $m_J$, is consequently
calculated.

\begin{center}
\begin{figure}[h]
\includegraphics[clip,width=1\linewidth]{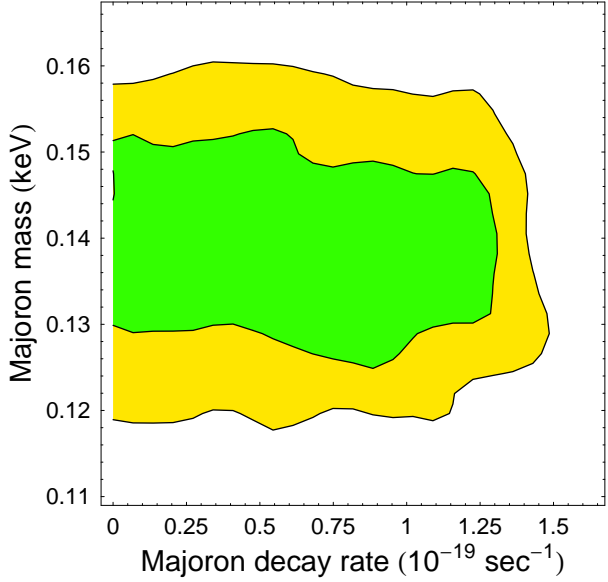}
\caption{(Color online) Contours of the 68\% (green/dark) and 95\% (yellow/light)
  confidence regions in the $(\Gamma_J,\,m_J)$ plane.}
\label{fig:cmb-mjvstau}
\end{figure}
\end{center}

We show our result in Fig.  ~\ref{fig:cmb-mjvstau}, where we give the
68\% and 95\% confidence contours in the $(m_J,\,\Gamma)$ plane, for
the case $\beta=1$, i.e., thermal majoron production and $N_D=427/4$.
We note that these parameters are not degenerate one with the other,
so the respective constraints are independent. Similarly, we find no
degeneracy between $\Gamma$ and the five standard parameters. The
marginalized 1-dimensional limits for $\Gamma$ and $m_J$ are:
\begin{eqnarray}
\Gamma < 1.3\times 10^{-19} \mathrm{sec}^{-1}\\
0.12\, \keV < m_J < 0.17\, \keV
\end{eqnarray}
Expressed in terms of the majoron lifetime our result implies $\tau >
250\, \mathrm{Gyr}$, nearly a factor 20 improvement with respect to
the naive limit $\tau~>~t_0~\simeq~14\,~\mathrm{Gyr}$, illustrating
the power of CMB observations in constraining particle physics
scenarios.
%
Let us comment on the possibility that $\beta\neq 1$. From eq.
\ref{eq:omj}, it can be seen that this amounts to the transformation
$m_J\to \beta m_J$. For example, as we have already pointed out, $N_D$ 
can be as large as $427/2\simeq 200$, so that
the above limit would read $0.24\,\keV<m_J<0.34\,\keV$. In general, if
we allow for the possibility of extra degrees of freedom in the early
Universe, we always have $\beta<1$ and then
$m_J>0.12\,\keV~. $	
If instead majorons are produced non-thermally, one will in general
have $\beta>1$.


Recent analysis of the Lyman-$\alpha$ forest data \cite{Seljak:2006qw} suggest
that the mass of the warm dark matter particle should be larger than at least 
1 keV (maybe even an order of magnitude more), the exact result depending on the candidate under consideration, on the
data used and on the analysis pipeline. Taken at face value, these results, combined
with the limits we obtain from the CMB, seem to exclude the
majoron as a viable dark matter candidate. However, care should be taken
in naively applying this results to the model presented here. First, the limits
on the mass of the WDM particle have been obtained in the case of a
stable particle. The effect of the decay on the growth of density fluctuations should be
taken into account to reliably compare the predicted matter power spectrum to
the observations. Second, the results of the Lyman-$\alpha$ forest analysis 
actually depend on the phase space distribution of the particles at the time of decoupling, 
and then ultimately on the production mechanism. On the contrary, our results, when quoted
in terms of the quantity $\beta m_J$, are completely independent on the production
mechanism. This is due to the fact that, as we commented above, it is a good approximation
to consider that the thermal velocities of majorons are negligible, as long as only
the CMB is concerned. This means that the particle mass never enter directly in the 
perturbation equations; instead it only enters indirectly through the background quantity 
$\Omega_J\propto\beta m_J$, and this should be regarded as the quantity that
is really constrained by CMB observations. A production mechanism resulting in a ``sub-thermal''
(i.e., $\beta<1$) majoron abundance, will result in the same dark matter energy density
being shared between a smaller number of particles, and then in a larger particle mass.
The same result of a larger mass for fixed $\Omega_J$ can be achieved if the number
of quantum degrees of freedom at decoupling is substantially larger than its standard model value
of 106.75, for example in theories with larger gauge groups and representations. Finally,
it should be reminded that there is still no consensus on whether the Lyman-$\alpha$ data
can be considered fully reliable, due the various systematics that are involved in
the analysis pipeline.


\paragraph{Particle physics model} We now briefly comment on the particle physics model. The simplest
possibility is that neutrino masses arise {\it a la
  seesaw}~\cite{Valle:2006vb}.  In the basis $\nu, \nu^c$ (where $\nu$
denote ordinary neutrinos, while $\nu^c$ are the \21 singlet
``right-handed'' neutrinos) the full neutrino mass matrix is given as
\begin{equation}
\label{ss-matrix-123} {\mathcal M_\nu} = \left(\begin{array}{cc}
    Y_3 v_3 & Y_\nu v_2 \\
    {Y_\nu}^{T} v_2  & Y_1 v_1 \\
\end{array}\right) 
\end{equation}
and involves, in addition to the singlet, also a Higgs triplet
contribution~\cite{schechter:1980gr} whose vacuum expectation value
obeys a ``vev seesaw'' relation of the type $v_3 v_1 \sim v_2^2$.
The Higgs potential combines spontaneous breaking of lepton number and
of the electroweak symmetry. The properties of the seesaw majoron and
its couplings follow from the symmetry properties of the potential and
were extensively discussed in \cite{schechter:1982cv}.
Here we assume, in addition, that quantum gravity
effects~\cite{Coleman:1988tj} produce non-renormalizable Planck-mass
suppressed terms which explicitly break the global lepton number
symmetry and provide the majoron mass, which we can not reliably
compute, but we assume that it lies in the cosmologically
interesting keV range.

In all of such models the majoron interacts mainly with neutrinos,
proportionally to their mass~\cite{schechter:1982cv}, leading to
\begin{equation}
  \label{eq:jj}
\tau  (J \to \nu \nu) \approx \frac{16\pi}{ m_J} \frac{v_{1}^2}{m_\nu^2}.
\end{equation}
The limits obtained above from the WMAP data can be used to roughly
constrain the lepton number breaking scale as $v_1^2 \gtrsim
3\times\left( 10^6\,\GeV \right)^2$, for $m_\nu \simeq 1\eV$.

\paragraph{Future perspectives} 
The massive majoron has also a sub-leading radiative decay mode, $J \to
\gamma \gamma$, making our DDM scenario potentially testable through
studies of the diffuse photon spectrum in the far ultra violet.  A
more extended investigation of these schemes will be presented
elsewhere including other cosmological data such as the large scale
structure data from the Sloan Digital Sky Survey (SDSS)~\cite{prepa}.
In contrast, we do not expect the data from upcoming CMB experiments
like Planck to substantially improve our bounds on the majoron decay
rate, since they mainly affect the large angular scales where the
error bars have already reached the limit given by cosmic variance.
We also note that direct detection of a keV majoron is possible in a
 suitable underground experiment \cite{Bernabei:2005ca}.
\paragraph{Note added in the arXiv version} After we sent this paper for publication,
another work appeared, studying the possibility that the entropy production
 and bulk viscosity associated to the late decay of the dark matter particle could explain the
observed acceleration of the Universe \cite{Mathews:2008hk}. However
the decay rate needed to accomplish this seems to be too large 
with respect to the limit found in the present analysis of the CMB data.
 
\subsubsection{Acknowledgements} ML is currently
supported by INFN.  Part of the work reported here was done when ML was visiting
IFIC (Valencia), supported by a fellowship from the University of Rome ``La
Sapienza''. The author would like to acknowledge J.W.F. Valle for useful discussion,
and S. Pastor for computing help. ML would like
to acknowledge ICRANET for having supported his participation to the meeting.

 \def\baselinestretch{1}%


\end{document}